\documentclass[prl,aps,twocolumn,preprintnumbers,showpacs]{revtex4}
\pdfoutput=1

\usepackage{amsmath,amssymb,bm}
\usepackage{graphicx}
\usepackage{color}

\begin{document}
\title{Magnetoconductance of carbon nanotube p-n junctions}
\author{A.~V.~Andreev}
\affiliation{Department of Physics, University of Washington,
Seattle, Washington 98195-1560, USA}
\date{June 5, 2007}
\begin{abstract}
The magnetoconductance of p-n junctions formed in clean single wall
carbon nanotubes is studied in the noninteracting electron
approximation and perturbatively in electron-electron interaction,
in the geometry where a magnetic field is along the tube axis. For
long junctions the low temperature magnetoconductance is anomalously
large: the relative change in the conductance becomes of order unity
even when the flux through the tube is much smaller than the flux
quantum. The magnetoconductance is negative for metallic tubes. For
semiconducting and small gap tubes the magnetoconductance is
nonmonotonic; positive at small and negative at large fields.
\end{abstract}
 \maketitle

Magnetoconductance arises from the orbital and Zeeman coupling of
electrons to the external magnetic field, $H$. As long as the flux
through the crystalline unit cell is much smaller than the flux
quantum $\Phi_0=hc/e$ magnetotransport may be described in the
semiclassical approximation~\cite{Peierls} ignoring the band
structure changes due to the presence of $H$. In most crystals this
condition holds at all experimentally realizable fields. Recently
much attention was focused on magnetotransport properties of carbon
nanotube devices. Because of their large radius the magnetic field
affects the one-dimensional electron spectrum~\cite{Ando} even at
relatively weak fields. This leads to interesting magnetotransport
phenomena~\cite{Man,Kong2001,Liang2001,Ivchenko2002,Cao2004,Durkop,Cobden2005,Fedorov2007}.

In this paper we show that the magnetic field dependence of the
one-dimensional band structure results in a peculiar mechanism of
magnetoconductance of p-n junctions in carbon nanotubes. This
mechanism is relevant to the magnetoresistance of nominally undoped
metallic and small gap nanotubes placed on an insulating substrate.
In this case the long range disorder potential caused by charged
impurities in the substrate creates p- and n- regions in the tube,
and backscattering of electrons arises mainly from the gaps between
p- and n- regions, where the semiclassical description of electron
transport fails.

\begin{figure}[ptb]
\includegraphics[width=8.0cm]{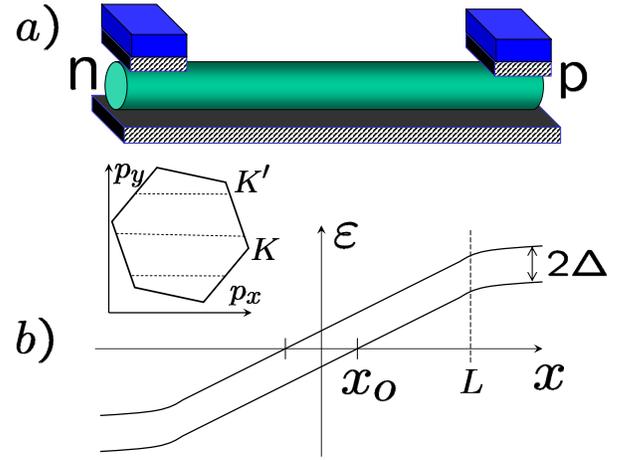}
\caption{a) Device sketch: a nanotube rests on an insulating
substrate. The n- and p- regions are created by biasing the top
gates. b) Band diagram of the device. Tilted solid lines represent
the bottom of the conduction and the top of the valence bands. The
width of the center region is $2L$. The width of the classically
forbidden region is $2x_o=2\Delta/eE$. Inset: one dimensional
spectrum is obtained by dissecting the graphene Brillouin zone by
the $p_y=0$ line (dashed line). The low energy states lie in the
vicinity of $K$ and $K'$ points. } \label{fig:setup}
\end{figure}

We study the magnetoconductance of a p-n junction formed in a clean
single wall carbon nanotube for magnetic fields parallel to the tube
axis.   The device is depicted in Fig.~\ref{fig:setup} $a)$. The p-
and n- regions  can be formed by appropriately biasing the top
gates.  Such devices were recently used to measure~\cite{Ilani2006}
thermodynamic properties of electron liquid in carbon nanotubes. It
is shown below that for realistic device parameters similar to those
of Ref.~\onlinecite{Ilani2006} the low temperature
magnetoconductance becomes of order unity while the flux $\Phi$
through the tube cross-section is much smaller than $\Phi_0$.

Before proceeding to detailed calculations let us qualitatively
discuss the origin of strong magnetoconductance. We choose the $x$-
and $y$- axes to be respectively along the tube axis and along its
circumference. The one-dimensional electron sub-band spectrum is
determined by intersections of the graphene Brillouin zone with the
$k_y=(\Phi/\Phi_0 +m)/R$ lines, see the inset in
Fig.~\ref{fig:setup} $b)$. Here $R$ is the tube radius and $m$ an
integer. The electron spectrum near the $K$ and $K'$ points becomes
$\pm \sqrt{\Delta^2+(\hbar \textsl{v} p_x)^2}$, where
$\textsl{v}\approx 8\times 10^5 \, m/s$ is the electron velocity in
graphene, and $\Delta$ is half the energy gap between the valence
and conduction sub-bands. In the center region between the p- and n-
banks the external potential is assumed to be approximately linear,
$U(x)=e E x$, where $e$ is the electron charge and $E$ the electric
field. The tilting of electron energy bands in the external
potential produces a spatial region of width $2 x_0= 2\Delta/eE$,
where electron motion is classically forbidden, see
Fig.~\ref{fig:setup} $b)$. The device conductance is governed by the
Landau-Zener tunneling across this region. With exponential accuracy
the tunneling probability may be found  in the WKB approximation.
The electron momentum $p_x$ along the tube depends on the position
as $p_x = \sqrt{(eEx-\epsilon)^2-\Delta^2 }/\textsl{v}$. This gives
the transmission probability ${\cal T}= \exp \left(\frac{2}{\hbar}
\mathrm{Im}\int p_x dx \right)= \exp[-\pi \Delta^2/(\hbar \textsl{v}
eE)]$. The rigorous calculation given below shows that the
pre-exponential factor is equal to unity. The magnetoconductance
arises from the flux dependence of the band gap~\cite{Ando},
$\Delta=\Delta_0 \pm \hbar \textsl{v}\Phi/R\Phi_0$, where $\Delta_0$
is half-the band gap at zero flux and the $\pm$ sign corresponds to
the different valleys, $K$ and $K'$. Accounting for electron spin
and the two valleys and neglecting the Zeeman splitting one obtains
for the device conductance
\begin{equation}\label{eq:conductance_0}
    G_{0}=\frac{2e^2}{h}\sum_{j=1,2}
    \exp\left[-\frac{\pi \hbar \textsl{v}}{  e E R^2}
    \left(\frac{\Delta_0 R }{\hbar \textsl{v}} +
    \frac{(-1)^j \Phi}{\Phi_0} \right)^2
    \right].
\end{equation}
If the electric field is not too strong, $eER \ll \hbar \textsl{v}/R
$, the magnetoconductance becomes of order unity while the flux is
still small, $\Phi \ll \Phi_0$. In the case of metallic tubes,
$\Delta_0=0$, the magnetoconductance is negative. For semiconducting
and small gap tubes the magnetoconductance is nonmonotonic; positive
at small fields and negative at large ones. The conductance maximum
is attained at $\Phi_{max}\approx\Phi_0 \Delta_0 R/\hbar
\textsl{v}$. For semiconducting tubes, $\Delta_0\approx\hbar
\textsl{v} /3R$, this gives $\Phi_{max}\approx\Phi_0/3$. For small
gap tubes the zero flux gap arises only due to curvature effects and
is rather small, $\Delta_0\ll \hbar \textsl{v}/R$. In this case the
conductance maximum is achieved at $\Phi_{max} \ll \Phi_0$. A
nonmonotonic magnetoresistance was recently observed in
Ref.~\onlinecite{Fedorov2007} in nanotube devices with a different
geometry.

In the noninteracting electron approximation the magnetoconductance
only weakly depends on the temperature $T$ as long as the latter is
smaller than the Fermi energy in the banks. In this regime the
energies of electrons participating in transport lie in the narrow
band of width $T$ around the chemical potential. In this energy
range deviations of electric potential from the linear form
$U(x)\approx eE x$ are negligible. This results in
energy-independent transmission coefficient and thus
temperature-independent conductance.

In the presence of electron-electron and electron-phonon
interactions electrons can be transferred between the p- and n-
regions at finite temperature by thermal activation. At  $T \gg
|\Delta-\Delta_0|=\hbar \textsl{v}\,\Phi/R\Phi_0$ the rate of
inelastic processes is practically independent of the magnetic field
and magnetoconductance arises mainly from the tunneling mechanism
discussed above. In this regime inelastic transfers shunt the
tunneling mechanism and suppress magnetoconductance. A crude
estimate of the characteristic temperature $T^*$, above which the
magnetoconductance suppression becomes significant, can be obtained
in the tunneling regime by equating the activation rate,
$\sim\int_0^\infty\frac{dx}{\ell_{in}}\exp[-(\Delta
+eEx)/T]=\frac{T}{eE\ell_{in}}\exp(-\Delta/T)$, with $\ell_{in}$
being the inelastic mean free path, to the tunneling rate, $\sim
\exp\left(-\frac{\pi \Delta^2}{eE\hbar \textsl{v}} \right)$.
According to this estimate the noninteracting electron result,
Eq.~(\ref{eq:conductance_0}) provides a good description of the
conductance for  $T< eE R$.

\begin{figure}[ptb]
\includegraphics[width=8.5cm]{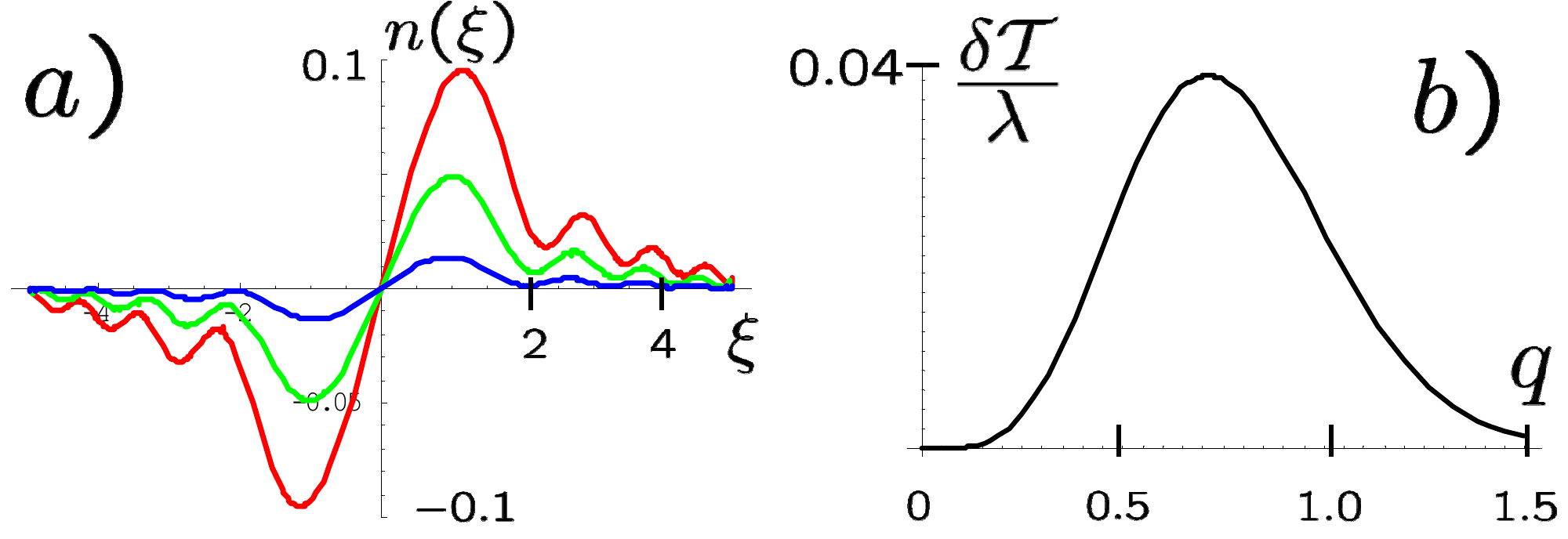}
\caption{a) Dimensionless electron density as a function of $\xi$
for $q=0.3$, $q=0.6$, and $q=0.9$. b) The ratio of the correction to
the transmission coefficient to the interaction constant $\lambda$,
Eq.~(\ref{eq:tT}), as a function of $q$.  } \label{fig:friedel}
\end{figure}

At zero temperature the finite reflection amplitude at the p-n
contact leads to the appearance of Friedel oscillations in the
electron density. The additional scattering of electrons from the
Friedel oscillations in the presence of electron-electron
interactions gives a correction~\cite{Yue1994} to the noninteracting
result for the device conductance, Eq.~(\ref{eq:conductance_0}).
This correction is evaluated below to first order in
electron-electron interaction. It is given by Eq.~(\ref{eq:tT}) and
is plotted in Fig.~\ref{fig:friedel} b). It remains small even if
the interaction constant, $e^2/\hbar \textsl{v}$ is of order unity.

The device conductance in the noninteracting electron approximation,
Eq.~(\ref{eq:conductance_0}), immediately follows from the results
of Cheianov and Falko~\cite{Cheianov2006} for a graphene p-n
junction that were obtained using transfer matrices. Below we
present a consideration in terms of wave functions that is more
convenient for the treatment of the interaction correction to
magnetoconductance. Electron eigenstates of energy $\varepsilon$
obey the Dirac equation, which by an appropriate basis choice can be
cast in the form
\[    [U(x)-\varepsilon -i\hbar \textsl{v} \,\sigma_z
    \partial_x+\Delta\sigma_y]\psi=0,
\]
where $\sigma_i$ are Pauli matrices. Introducing the dimensionless
coordinate $\xi=e E x/\sqrt{eE\hbar \textsl{v}}$, energy $\epsilon
=\varepsilon/\sqrt{eE\hbar \textsl{v}}$, and momenta
$q=\Delta/\sqrt{\textsl{v}\hbar eE}$ and
$k(\xi)=(U(\xi)-\varepsilon)/\sqrt{\textsl{v} \hbar eE}$ we rewrite
this equation as
\begin{equation}\label{eq:Sch}
    \left(
      \begin{array}{cc}
        k(\xi) -i \partial_\xi& -iq \\
        iq & k(\xi)+i \partial_\xi \\
      \end{array}
    \right)\left(
             \begin{array}{c}
               u \\
               v \\
             \end{array}
           \right)=0.
\end{equation}
The dimensionless momentum $k(\xi)$ changes from $-k_f$ at $\xi \to
- \infty$ to $+k_f$ at $\xi \to + \infty$, where $k_f\gg 1$ is the
dimensionless Fermi momentum in the banks. In the center region
between the banks the coordinate-dependent momentum $k(\xi)$ is
linear in $\xi$, $k(\xi)\approx \xi-\epsilon$, and the spinor
amplitudes $u$ and $v$ satisfy the differential equation,
\[
    (\partial_z^2 +a+z^2)f=0.
\]
Here $a=i-q^2$ for $u$, and $a=-i-q^2$ for $v$, and we introduced
the difference coordinate, $z=\xi-\epsilon$ . The independent
solutions of this equation are parabolic cylinder
functions~\cite{Morse} that can be expressed in terms of the
confluent hypergeometric function $F(\alpha, \gamma, z)$,
\begin{subequations}
\label{eq:even_odd}
\begin{eqnarray}
  f_e(z) &=& \exp\left(-i\frac{z^2}{2}\right)
  F\left(\frac{1}{4} +\frac{ia}{4}, \frac{1}{2},iz^2 \right), \nonumber \\
  f_o(z) &=& z \exp\left(-i\frac{z^2}{2}\right)
  F\left(\frac{3}{4} +\frac{ia}{4}, \frac{3}{2},iz^2 \right). \nonumber
\end{eqnarray}
\end{subequations}

Two linearly independent solutions of Eq.~(\ref{eq:Sch}) are
\begin{equation}\label{eq:Psi_12}
    \psi_1=e^{-\frac{\pi q^2}{4}}
    \left(
      \begin{array}{c}
        u_e\\
        - u_o^* \\
      \end{array}
    \right), \quad
    \psi_2=e^{-\frac{\pi q^2}{4}}
    \left(
      \begin{array}{c}
         -u_o\\
         u_e^*\\
      \end{array}
    \right),
\end{equation}
where
\begin{subequations}
\label{eq:u_even_odd}
\begin{eqnarray}
  u_e(z) &=& \exp\left(-i\frac{z^2}{2}\right)
  F\left(-i\frac{q^2}{4}, \frac{1}{2},iz^2 \right), \\
  u_o(z) &=& q\, z \exp\left(-i\frac{z^2}{2}\right)
  F\left(\frac{1}{2} -i\frac{q^2}{4}, \frac{3}{2},iz^2 \right).
\end{eqnarray}
\end{subequations}

Equation (\ref{eq:Sch}) conserves the current along the tube axis,
$I_x=\psi^\dagger\sigma_z\psi$. From the form of the current
operator it is clear that the top/bottom components of the
pseudo-spinor $\psi$ represent the amplitudes of the right-/left-
moving waves. Thus the scattering states incident from the left,
$\psi_L$, and right, $\psi_R$, can be found by requiring that the
bottom/top component of the spinor vanish at $\xi\to \pm \infty$.
Using Eqs.~(\ref{eq:Psi_12}) and (\ref{eq:u_even_odd}) and the large
distance asymptotics of the confluent hypergeometric
function~\cite{Morse},
\begin{equation}\label{eq:hyper_asympt}
    F(\alpha,\gamma,z\to \infty) \approx
    \frac{\Gamma(\gamma)}{\Gamma(\gamma-\alpha)}(-z)^{-\alpha} +
    \frac{\Gamma(\gamma)}{\Gamma(\alpha)}e^z z^{\alpha-\gamma},
\end{equation}
we obtain for the scattering states incident from the right and
left,
\begin{subequations}
\label{eq:psi_RL}
\begin{eqnarray}\label{eq:spinor_incident_left}
    \psi_L&=&   \left(
    \begin{array}{c}
    u_L \\
    v_L\\
    \end{array}
    \right)
    =\frac{\psi_1+\alpha\,\psi_2}{\sqrt{1+|\alpha|^2}}
    ,  \\
\psi_R&=&
 \left(
    \begin{array}{c}
    u_R \\
    v_R\\
    \end{array}
    \right)=
\frac{\psi_2-\alpha^*\,\psi_1}{\sqrt{1+|\alpha|^2}}.
\end{eqnarray}
\end{subequations}
Here $\alpha$ is given by
\begin{equation}\label{eq:alpha}
    \alpha=e^{-i\frac{\pi}{4}}\,\frac{q}{2}\, \frac{\Gamma
    \left(\frac{1}{2}-i\frac{q^2}{4}\right) }{
    \Gamma \left(1-i\frac{q^2}{4}\right)}, \quad |\alpha|^2=\tanh\frac{\pi
    q^2}{4}.
\end{equation}

The $z \to \pm \infty$ asymptotics of the right-moving wave (the top
spinor component) in the scattering states $\psi_L$ and $\psi_R$
are;
\begin{subequations}
\label{eq:incident_asympt}
\begin{eqnarray}
\label{eq:incident_asympt_a}
  u_L(z) &\approx&
    \frac{\sqrt{\pi} e^{- \frac{\pi q^2}{8}}
    [1-\mathrm{sgn}(z)|\alpha|^2]\, |z|^{i\frac{q^2}{2}}e^{-i\frac{z^2}{2}}}{
    \sqrt{1+|\alpha|^2}\Gamma \left(\frac{1}{2} +
  i \frac{q^2}{4}\right) },
 \\
    u_R(z) &\approx& -\frac{q}{2}\frac{\sqrt{ i \pi} e^{ -\frac{\pi q^2}{8}}
    [1+\mathrm{sgn}(z)]
    \, |z|^{i\frac{q^2}{2}}
   e^{-i\frac{z^2}{2}} }{\sqrt{1+|\alpha|^2}\Gamma \left(1 +
  i \frac{q^2}{4}\right) } .
   \label{eq:incident_asympt_b}
\end{eqnarray}
\end{subequations}
From Eq.~(\ref{eq:incident_asympt_a}) one finds the transmission
amplitude,
\[
    t_0\equiv\lim_{\xi \to +\infty
    }\frac{u_L(\xi)}{u_L(-\xi)}=\frac{1-|\alpha|^2}{1+|\alpha|^2}=\exp\left(-\frac{\pi
    q^2}{2}\right).
\]
This gives the transmission coefficient is ${\cal T}_0= \exp(-\pi
q^2)$, in agreement with Ref.~\onlinecite{Cheianov2006}. Expressing
$q$ in terms of the system parameters using expressions presented in
the text above Eq.(\ref{eq:Sch}), and taking into account the
magnetic field dependence of the energy gap one obtains the device
conductance, Eq.~(\ref{eq:conductance_0}).

Next we evaluate the first interaction correction to
Eq.~(\ref{eq:conductance_0}) at zero temperature. To first order in
interaction the correction to the device conductance can be obtained
by considering the change in the transmission amplitude for a
particle at the Fermi level that arises from the additional
scattering from the Hartree-Fock potential induced by the electron
density~\cite{Yue1994}. The induced Hartree-Fock potential has two
qualitatively different effects on the transmission amplitude: i) By
enhancing the effective electric field inside the classically
forbidden region it increases the tunneling amplitude, and ii) It
causes additional backscattering from the Friedel oscillations in
the classically allowed region. The analysis below shows that
repulsive interaction increases the transmission amplitude.

Using Eqs.~(\ref{eq:u_even_odd}) and (\ref{eq:hyper_asympt}) it is
easy to show that the spinor wave functions in Eq.~(\ref{eq:Psi_12})
are normalized to a $\delta$-function of the dimensionless energy,
\begin{equation}\label{eq:Psi_12_normalized}
   \int_{-\infty}^{\infty}
   d\xi\psi^\dagger_i(\xi-\epsilon)\psi_j(\xi-\epsilon')
    =2\pi \delta (\epsilon-\epsilon')\delta_{ij}.
\end{equation}
Therefore the electron density, upon subtraction of the uniform ion
background, is
\begin{equation}\label{eq:density}
    n(\xi)= \sum_{i=1,2}
    \int_{-\infty}^{\infty} \frac{ \mathrm{sign}(-\epsilon)
    d\epsilon}{4\pi}\psi_i^\dagger(\xi-\epsilon) \psi_i(\xi - \epsilon).
\end{equation}
In this equation the energy is measured from the Fermi level and the
electron hole symmetry of the problem was used. Plots of electron
density, Eq.~(\ref{eq:density}), for different values of $q$ are
presented in Fig.~\ref{fig:friedel} a). The Friedel oscillations in
charge density appear only at finite refection amplitude and fall
off as $1/\xi^2$. The extra power of $1/\xi$ in comparison with the
usual one-dimensional Fermi gas arises from the linearly growing
external potential. Because of the fast decay of the oscillations
the correction to the transmission amplitude is free from infrared
divergences and arises from distances $x \sim \sqrt{\hbar \textsl{v}
/eE}$.

Below we consider the case of a short range interaction. This should
be a reasonable approximation because the Fourier transform of the
Coulomb interaction depends on the transferred momentum only
logarithmically. Since the characteristic scattering momentum is
$\sqrt{\hbar e E/\textsl{v}}$ the dimensionless interaction constant
can be estimated as $\lambda \sim \frac{1}{1+\kappa}\frac{e^2}{\hbar
\textsl{v}}\ln\frac{ eE R^2}{\hbar \textsl{v}}$, where $\kappa$ is
the dielectric constant of the substrate. We restrict the
consideration to metallic tubes, for which electron spectra in the
presence of the flux remain degenerate in the two valleys. Then the
Hartree-Fock potential is $V(\xi)= 3\lambda \, n(\xi)$, where the
factor $3=4-1$ arises from the spin and valley degeneracy.

In the presence of the perturbation potential the wave incident from
the left at the Fermi level, $\epsilon=0$, can be written as $
\psi_L(\xi) +\chi (\xi)$. The correction, $\chi (\xi)$, to the wave
function satisfies the equation $[\hat{H}_0+V(\xi)]\chi
(\xi)=-V(\xi)\psi_L(\xi)$ with $\hat{H}_0=\xi -i\partial_\xi
\sigma_z +q\sigma_y$ being the unperturbed Hamiltonian. To first
order in perturbation the solution of this equation is
\begin{equation}\label{eq:chi_GF}
    \chi (\xi)=\int d\xi' G^R(\xi, \xi')V(\xi')\psi_L(\xi'),
\end{equation}
where $G^R(\xi, \xi')=(\hat{H}_0+i\eta)^{-1}$ is the Green's
function that can be expressed in terms of the spinors in
Eq.~(\ref{eq:Psi_12}),
\begin{equation}\label{eq:GF}
    G^R(\xi,
    \xi')=
    \sum_{i=1,2}\int_{-\infty}^{\infty}\frac{d\epsilon}{2\pi}
    \frac{\psi_i(\xi-\epsilon)
    \psi^\dagger_i(\xi'-\epsilon)}{\epsilon+i\eta}.
\end{equation}

In order to find the correction to the transmission amplitude one
needs only the large distance asymptotics of $\chi(\xi)$. Therefore
only the on-shell part of the Green's function will contribute to
Eq.~(\ref{eq:chi_GF}) at $\xi \to \infty$,
\begin{equation}\label{eq:chi_infty}
    \chi (\xi)=-\frac{i}{2}\sum_i\psi_i(\xi)
    \int d\xi' \psi^\dagger_i(\xi')V(\xi')\psi_L(\xi').
\end{equation}
Since the wave functions $\psi_{1,2}$ are related to the scattering
states $\psi_{L,R}$ by the unitary transformation
Eq.~(\ref{eq:psi_RL}) the sum over $i$ in Eq.~(\ref{eq:chi_infty})
can be understood to run over $i=L,R$. Next we notice that for $i=L$
the integral in Eq.~(\ref{eq:chi_infty}) is purely real. Therefore
this term will only change the phase of the transmission amplitude,
but not its modulus. Thus to compute the first correction to the
transmission amplitude we may replace $\chi(\xi)$ by
\[    \tilde{\chi} (\xi)=-\frac{i}{2}\psi_R(\xi)
    \int d\xi' \psi^\dagger_R(\xi')V(\xi')\psi_L(\xi').
\]
The transmission amplitude can be found from the equation
$t=t_0+\lim_{\xi\to +\infty}\frac{(1, 0)\cdot
\tilde{\chi}(\xi)}{u_L(-\xi)}$. Using
Eq.~(\ref{eq:incident_asympt_b}) we obtain for the correction
$\delta {\cal T}$ to the transmission coefficient,
\begin{equation}\label{eq:tT}
    \frac{\delta {\cal T}}{\lambda}=
    -\frac{3 e^{-\frac{\pi q^2}{2}}}{1+|\alpha|^2}
    \int d\xi'
    n(\xi')\mathrm{Im}[\alpha^*\psi^\dagger_R(\xi')\psi_L(\xi')].
\end{equation}
It may be evaluated using Eqs.~(\ref{eq:psi_RL}), (\ref{eq:Psi_12}),
(\ref{eq:u_even_odd}) and (\ref{eq:density}), and is plotted as a
function of $q$ in Fig.~\ref{fig:friedel} b). At weak reflection, $q
< 1$, the correction to transmission coefficient is small even if
the interaction constant is of order unity. This is because the
Friedel oscillations amplitude is proportional to the reflection
amplitude. Thus the noninteracting result,
Eq.~(\ref{eq:conductance_0}), provides a good description for the
low temperature conductance of metallic devices. In the tunneling
regime, $q\gg 1$, the relative correction to the transmission
amplitude is expected to be strong because of the exponential
dependence of the tunneling amplitude on the effective electric
field. Therefore the method described above becomes inapplicable.

In summary, the low temperature magnetoconductance of p-n junctions
in clean single wall carbon nanotubes was studied in the geometry
where the magnetic field is along the tube axis. For weak
band-tilting field $E \ll \hbar \textsl{v}/e R^2$ the
magnetoconductance of long, $L\gg R$, junctions becomes of order
unity while the flux through the tube is much smaller than the flux
quantum. In the noninteracting electron approximation the device
conductance is given by Eq.~(\ref{eq:conductance_0}). The
magnetoconductance is positive for metallic tubes and nonmonotonic
for semiconducting and small gap tubes. The interaction correction
to the zero temperature magnetoconductance was studied to first
order in perturbation theory. It arises due to the change in the
effective electric field in the gap between the p- and n- regions
and due to the scattering from the Friedel oscillations. In contrast
to the one-dimensional Fermi gas, the Friedel oscillations in the
present geometry fall off as $1/x^2$, which leads to the absence of
infrared divergence in the correction to the tunneling amplitude.
The net correction to the tunneling probability is positive. It is
given by Eq.~(\ref{eq:tT}) and is plotted in Fig.~\ref{fig:friedel}
b). If the reflection coefficient is not too strong the
noninteracting electron result, Eq.~(\ref{eq:conductance_0}), is
rather accurate even if the coupling constant is of order unity. At
finite temperatures, in addition to tunneling across the classically
forbidden region electrons can be transferred between p- and n-
regions by being promoted across the band gap due to inelastic
electron-electron and electron phonon scattering. The estimate of
activation transfer rate shows that tunneling processes described by
Eq.~(\ref{eq:conductance_0}) dominate the transport for $T< eER$.

I would like to thank D. Cobden, E. Mishchenko, T.~D. Son and B.
Spivak and for useful discussions.

\end{document}